\documentclass[amsmath,amssymb,aps,nofootinbib,prd,preprint,tightenlines]{revtex4-2}

\usepackage{bm}
\usepackage[colorlinks,citecolor=blue,urlcolor=blue,linkcolor=red]{hyperref}
\usepackage[mathscr]{euscript}
\usepackage{graphicx}
\usepackage{multirow,slashed}
\usepackage{xcolor}
\usepackage{booktabs}

\begin{document}
	
	\baselineskip=17pt \parskip=5pt
	
	\title{Orbital Stability Study of the Taiji Space Gravitational Wave Detector}
	\author{Yu-Yang Zhang\,$^{1,2,3}$\footnote[1]{zhangyuyang21@mails.ucas.ac.cn}, Geng Li\,$^{2,4}$\footnote[2]{ligeng@ucas.ac.cn (Corresponding Author)} and Bo Wen\,$^{1,2,5}$\footnote[3]{wenbo21@mails.ucas.ac.cn}\\}
	\affiliation{
		$^{1}$ School of Fundamental Physics and Mathematical Sciences, Hangzhou Institute for Advanced Study, UCAS, Hangzhou 310024, China\\
		$^{2}$ University of Chinese Academy of Sciences, Beijing 100049, China\\
		$^{3}$ Institute of Theoretical Physics, Chinese Academy of Sciences, Beijing 100190, China\\
		$^{4}$ Institute of High Energy Physics, Chinese Academy of Sciences, Beijing 100049, China\\
		$^{5}$ National Space Science Center, Chinese Academy of Sciences, Beijing 100190, China\\}
	
	\begin{abstract}
		
		Space-based gravitational wave detection is extremely sensitive to disturbances. The Keplerian configuration cannot accurately reflect the variations in spacecraft configuration. Planetary gravitational disturbances are one of the main sources. Numerical simulation is an effective method to investigate the impact of perturbation on spacecraft orbits. This study shows that, in the context of the Taiji project, Earth's gravity is an essential factor in the change in heliocentric formation configuration, contributing to the relative acceleration between spacecrafts in the order of $\mathcal O(10^{-6})\,{\rm m\cdot s^{-2}}$. Considering 00:00:00 on 27 October 2032 as the initial orbiting moment, under the influence of Earth's gravitational perturbation, the maximum relative change in armlengths and variation rates of armlengths for Taiji is $1.6\times 10^{5}\,{\rm km}$, $32\,{\rm m\cdot s^{-1}}$, respectively, compared with the unperturbed Keplerian orbit. Additionally, by considering the gravitational perturbations of Venus and Jupiter, the armlength and relative velocity for Taiji are reduced by $16.01\%$ and $17.45\%$, respectively, compared with when only considering that of Earth. The maximum amplitude of the formation motion indicator changes with the orbit entry time. Results show that the relative velocity increase between the spacecrafts is minimal when the initial orbital moment occurs in July. Moreover, the numerical simulation results are inconsistent when using different ephemerides. The differences between ephemerides DE440 and DE430 are smaller than those between DE440 and DE421. 
		
	\end{abstract}
	
	\maketitle
	
	\newpage
	
	
	\section{Introduction}
	
	On 11 February  2016, the Laser Interference Gravitational-Wave Observatory (LIGO)\,\cite{GW150914} announced the direct observation of gravitational wave signals, opening a new era of physics\,\cite{GW170608,GW170814}. Compared with ground-based gravitational wave detection, space-based detection can address the limitations of ground-based noise\,\cite{GWO1,GWO2,GWO3} and interferometer scale, enabling the detection of gravitational waves in the medium- and low-frequency bands\,\cite{LISA_conclusion,JOFFRE20213868,LISA_formation2}. The success of the Laser Interferometer Space Antenna Pathfinder (LPF) validates the feasibility of gravitational wave detection in space\,\cite{LISAPathfinder1,LISAPathfinder2,LISAPathfinder3,LISAPathfinder4}. The subsequent detection plan, the Laser Interferometer Space Antenna (LISA) project, is currently under pre-research and simulation, with a targeted launch date of around 2035\,\cite{bayle2022overview}.
	
	In 2016, the Chinese Academy of Sciences proposed the Taiji Space Gravitational Wave Detection Program\,\cite{Hu:2017mde,WYL}. The program aims to use Taiji spacecrafts to form a formation of three spacecrafts in Earth-like orbits around the Sun, describing an equilateral triangle with a side length of $3$ million km\,\cite{taiji2}. These three spacecrafts would be distributed approximately $20^{\circ}$ in front of (or behind) the Earth, and the angle between the spacecraft formation plane and the ecliptic plane would be $60^{\circ}$\,\cite{Luo:2019zal,taijiangle20}. 
	
	{Space-based gravitational wave detection detects gravitational waves through the use of laser interferometry\,\cite{LISA:2017pwj}. As a gravitational wave passes through, it weakly distorts space, causing a small change in the length of the interferometer arms. By measuring the interference effect of the laser beam in the interferometer, this small change in arm length can be detected, and a gravitational wave signal is obtained. }In space-based gravitational wave detection, the length of the laser interference arm changes with time\,\cite{armlength}. The instability of the laser frequency is one of the main sources of noise. Time-delay interferometry (TDI) technology\,\cite{TDI,Otto2015TimedelayIS} is proposed to reduce this measurement noise. The implementation of TDI requires precise spacecraft absolute range measurement and real-time communication. Thus, the stability of spacecraft orbits should be investigated. Spacecrafts operate in a complicated gravitational environment. The orbit calculations must consider a variety of disturbances, such as planetary gravity, planetary tidal force, lunar gravity, solar light pressure, and post-Newtonian effects\,\cite{Pucacco_2010}. Planetary gravity is one of the dominant factors, especially the gravities of Earth, Venus, and Jupiter, which has a significant impact on orbital stability\,\cite{JOFFRE20213868,Halloin_2017}. 
	
	Furthermore, to meet the requirements for gravitational wave detection within targeted frequency bands, stringent constraints are imposed on the parameters governing the configuration of the Taiji formation\,\cite{Taiji_constraint}.
	\begin{itemize}
		\item[$\bullet$] To mitigate significant shifts within the sensitive frequency band, it is essential to limit the arm length of the Taiji constellation within a range of $3\pm 0.035$ million kilometers.
		\item[$\bullet$] The Taiji constellation does not need to keep a rigid geometry, but the change rate of armlength is limited to under $10\,{\rm m/s}$.
		\item[$\bullet$] Due to the design requirements of the optical system, the breathing angle of the Taiji constellation should be maintained at $60 \pm 1$ degrees.
	\end{itemize}
	
	To meet the aforementioned detection requirements, the configuration of the Taiji formation needs to maintain orbital stability. Gravitational perturbations from celestial bodies induce changes in spacecraft velocities, thereby altering their orbits. Over time, these variations may result in the armlength of spacecraft, breathing angle, and velocities exceeding the maximum acceptable limit. Once  this threshold is surpassed, we will be unable to effectively probe the target frequency band for the gravitational wave detection of Taiji. 
	Hence, assessing the impact of planetary gravitational perturbations on satellite formations is crucial for calculating experimental duration and facilitating subsequent orbit optimization~efforts.
	
	This study calculates the armlength, armlength variation rate, and breathing angle variation of the Taiji constellation under planetary gravitational perturbations. Employing ephemeris DE440 and based on the Keplerian orbit of Taiji, these variations under the influence of the gravities of Venus, Earth, and Jupiter were evaluated. Our analysis showed that the variations in the heliocentric formation configuration are primarily influenced by Earth's gravitational perturbations. Numerical simulations based on ephemeris DE440 show that the contribution of the Earth's gravity to the relative acceleration between spacecrafts is approximately $\mathcal O(10^{-6})\,{\rm m\cdot s^{-2}}$. Considering 00:00:00 on 23 October 2032 as the initial orbiting moment as an example, when the Earth's gravitational perturbation is determined, within six years the maximum amplitude of the relative variation in the armlength is $1.6\times10^5\,{\rm km}$, maximum amplitude of the relative variation in breathing angle is $3.2^{\circ}$, and maximum amplitude of the relative variation in the armlength variation rates is $32\,{\rm m\cdot s ^{-1}}$, compared with the unperturbed Keplerian configuration. By adding the gravitational perturbations of Venus and Jupiter, the armlength and relative velocity are reduced by $16.01\%$ and $17.45\%$, respectively, compared with when only the Earth's gravitational perturbation is considered. 
	Results showed that the maximum amplitude of Taiji constellation armlengths, armlength variation rate, and breathing angles varied with orbital entry moments. The maximum amplitude of the inter-spacecraft armlength change rate is the smallest when the initial orbital moment occurs in July. When the orbital entry moment is 00:00:00 on 1 July 2032, the maximum amplitude of the reduced armlengths is approximately $8\times10^4\,{\rm km}$, the maximum amplitude of the reduced breathing angles is approximately $1.6^{\circ}$, and the maximum armlength variation rate is approximately $18.1\,{\rm m\cdot s^{-1}}$ within six years.
	
	We show that by using different ephemerides for the calculations, the constellation armlengths deviate in the order of $\mathcal O(1)\,{\rm m}$, constellation armlength variation rate deviates in the order of $\mathcal O(10^{-7})\,{\rm m\cdot s^{-1}}$, and relative accelerations deviate in the order of $\mathcal O(10^{-13})\,{\rm m\cdot s^{-1}}$ within six years. The differences between ephemerides DE440 and DE430 are smaller than those between  DE440 and DE421.
	
	The remainder of the paper is presented as follows: 
	In Section\,\ref{sec2}, we derive the expressions for the positions and velocities of spacecrafts in the Keplerian configuration. The armlengths, armlength variation rate, inter-spacecraft relative acceleration, and breathing angles are determined. 
	In Section\,\ref{sec3}, we discuss the effect of planetary gravitational perturbations on the heliocentric formation configuration and the effect of different entering orbit moments. 
	In Section\,\ref{sec4}, we compare the calculation results using different ephemerides. 
	In Section\,\ref{sec5}, we present the conclusions. 
	
	\section{The Keplerian Orbital Configuration of Taiji}\label{sec2}
	
	\subsection{The Keplerian Orbit}
\begin{figure}[htbp]
		\centering
		\includegraphics[width=0.7\textwidth]{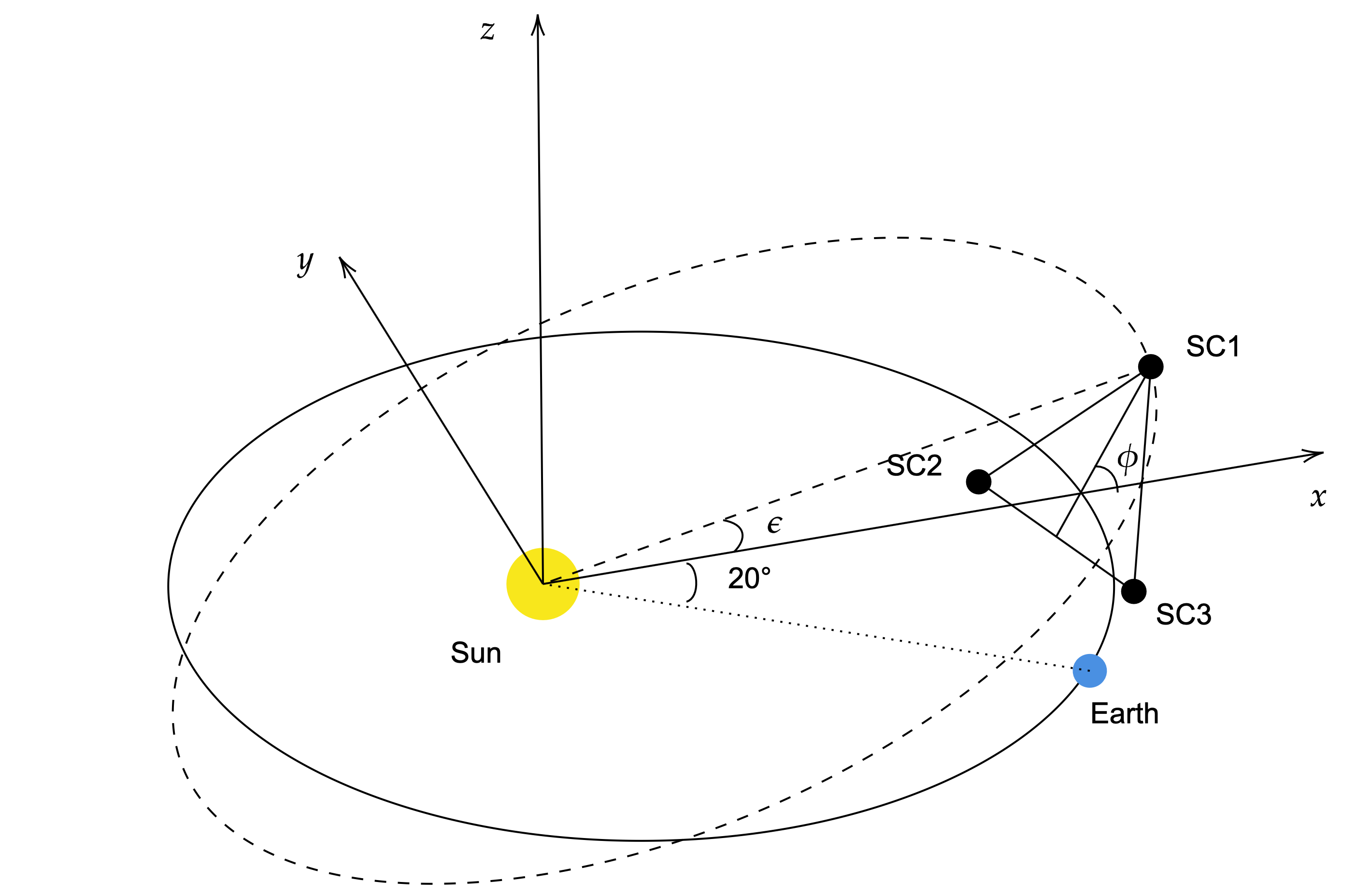}
		\caption{Taiji configuration and orbit.}
		\label{Fig: Taiji configuration and orbit.}
	\end{figure}

	Several orbitals have been designed for the Taiji heliocentric formation configuration\,\mbox{\cite{JOFFRE20213868,LISA_formation2,ruan2020lisa}}. We adopted the Keplerian orbital configuration model proposed in ref.\,\cite{Taiji_analysis} as part of the Taiji program preview, as shown in Figure\,\ref{Fig: Taiji configuration and orbit.}. 
	In the Keplerian configuration, the heliocentric reference system ECLIPJ2000 is chosen, the argument of periapsis is denoted as $\omega$, and the longitude of the ascending node is denoted as $\Omega$. When $t = 0$, spacecraft 1 (SC1) is at the highest point of the orbit and is located in the XZ plane. This can be associated with the determination of $\omega$ and $\Omega$ of SC1 as $270^{\circ}$\,\cite{formation}. As the three orbits have rotational symmetry about the Z-axis with a rotation angle of $120^{\circ}$, the $\omega$ values of spacecraft 2 (SC2) and spacecraft 3 (SC3) are the same as that of SC1 with $\omega = 270^{\circ}$. SC2 has $\Omega = 30^{\circ}$, and SC3 has $\Omega = 150^{\circ}$. Notably, the orbital semi-major axis of the three spacecrafts is $a$, eccentricity is $e$, and inclination is $\epsilon$. The spacecraft position vectors $\boldsymbol{r}_i = (x_i,y_i,z_i)\ (i = 1,2,3)$ and the spacecraft velocity vectors $\boldsymbol{v}_i = (v_{xi},v_{yi},v_{zi})\ (i = 1,2,3)$ can be expressed as follows: 
	\begin{equation}
		\label{Equ: Kepler position}
		\left\{\begin{aligned}		               
			x_i&=a(e+\cos{E}_i)\cos\varepsilon\cos\left[(i-1)\frac{2\pi}3\right]-a\sqrt{1-e^2}\sin{E}_i\sin\left[(i-1)\frac{2\pi}3\right],\\
			y_i&=a(e+\cos{E}_i)\cos\varepsilon \sin\left[(i-1)\frac{2\pi}3\right]+a\sqrt{1-e^2}\sin{E}_i\cos\left[(i-1)\frac{2\pi}3\right],\\
			z_i&=a(e+\cos{E}_i)\sin\varepsilon,
		\end{aligned}\right.
	\end{equation}
	and
	\begin{equation}\label{Equ: Kepler velocity}
		\left.\left\{\begin{aligned}
			v_{xi}&=-a(e+\sin{E}_i)\cos\varepsilon\cos\left[(i-1)\frac{2\pi}3\right]\frac{d{E}_i}{dt}-a\sqrt{1-e^2}\cos{E}_i\sin\left[(i-1)\frac{2\pi}3\right]\frac{d{E}_i}{dt},\\
			v_{yi}&=-a(e+\sin{E}_i)\cos\varepsilon \sin\left[(i-1)\frac{2\pi}3\right]\frac{d{E}_i}{dt}+a\sqrt{1-e^2}\cos{E}_i\cos\left[(i-1)\frac{2\pi}3\right]\frac{d{E}_i}{dt},\\
			v_{zi}&=-a(e+\sin{E}_i)\sin\varepsilon\frac{d{E}_i}{dt},
		\end{aligned}\right.\right.
	\end{equation}
	where $E_i$ is the eccentric anomaly, which can be obtained from the Kepler equation
	\begin{equation}\label{Equ: Kepler equation}
		{E}_i+e\sin{E}_i=\Bar{\omega} t-(i-1)\frac{2\pi}{3},
	\end{equation}
	where $\Bar{\omega}=\sqrt{\frac{G(m_s+m_i)}{a^3}}$ is the average angular velocity of the spacecraft.
	
	In the Taiji program, the half-length axis is $a = 1\,{\rm AU}$, the orbital eccentricity is $e\approx5.789\times10^{-3}$, and the orbital inclination $\varepsilon$ is given by Equation\,\eqref{Equ: epsilon}\,\cite{Taiji_analysis}
	\begin{equation}
		\label{Equ: epsilon}
		\cos\varepsilon =\frac{\sqrt{3}}3\frac{\sqrt{3}+2\alpha\cos\phi}{1+e},\quad \sin\varepsilon=\frac{\sqrt{3}}3\frac{2\alpha\sin\phi}{1+e},
	\end{equation}
	where $\phi$ is the angle between the plane of spacecraft formation and ecliptic plane, and $\alpha$ is a small parameter for the expansion, $\phi = \frac{\pi}{3} + \frac{5\sqrt{3}e}{8},\ \alpha = \frac{\sqrt{3}}2\left(\sqrt{e^2 + 2e + \cos^2\phi} - \cos\phi\right)$. The spacecraft position $\boldsymbol{r}_i$ and velocity $\boldsymbol{v}_i$ varied with time $t$, as expressed in the above equation.
	
	\subsection{Configuration Parameter}
	The armlength $l_{ij}$, armlength variation rate $v_{i}$, relative acceleration between the spacecrafts $\ddot l_{ij}$, and breathing angle $\beta_{i}$ of the constellation are as follows:
	\begin{equation}\label{Equ:armlength, velocity, breathing angle, and relative acceleration}
		\begin{aligned}
			l_{ij} = |\boldsymbol{r}_i - &\boldsymbol{r}_j|, \quad
			\dot{l}_{ij} = \frac{(\boldsymbol{\dot{r}}_i - \boldsymbol{\dot{r}}_j) \cdot \boldsymbol{l}_{ij}}{l_{ij}} = v_{ij}, \quad
			\beta_{i} = \arccos\left(\frac{{\boldsymbol{r}_i \cdot \boldsymbol{r}_j}}{{|\boldsymbol{r}_i| \cdot |\boldsymbol{r}_j|}}\right), \quad\\
			& \ddot{l}_{ij} = \frac{(\ddot{\boldsymbol{r}}_i - \ddot{\boldsymbol{r}}_j)(\boldsymbol{r}_i - \boldsymbol{r}_j) + (\dot{\boldsymbol{r}}_i - \dot{\boldsymbol{r}}_j)^2(1 - \cos{\theta}^2)}{l_{ij}},
		\end{aligned}
	\end{equation}
	where $\theta$ is the angle between vector $\dot{\boldsymbol{r}}_i - \dot{\boldsymbol{r}}_j$  and vector $\boldsymbol{r}_i - \boldsymbol{r}_j$ and $i,j = 1,2,3$.
	
\begin{figure}[htbp]
		\centering
		\includegraphics[width=0.95\textwidth]{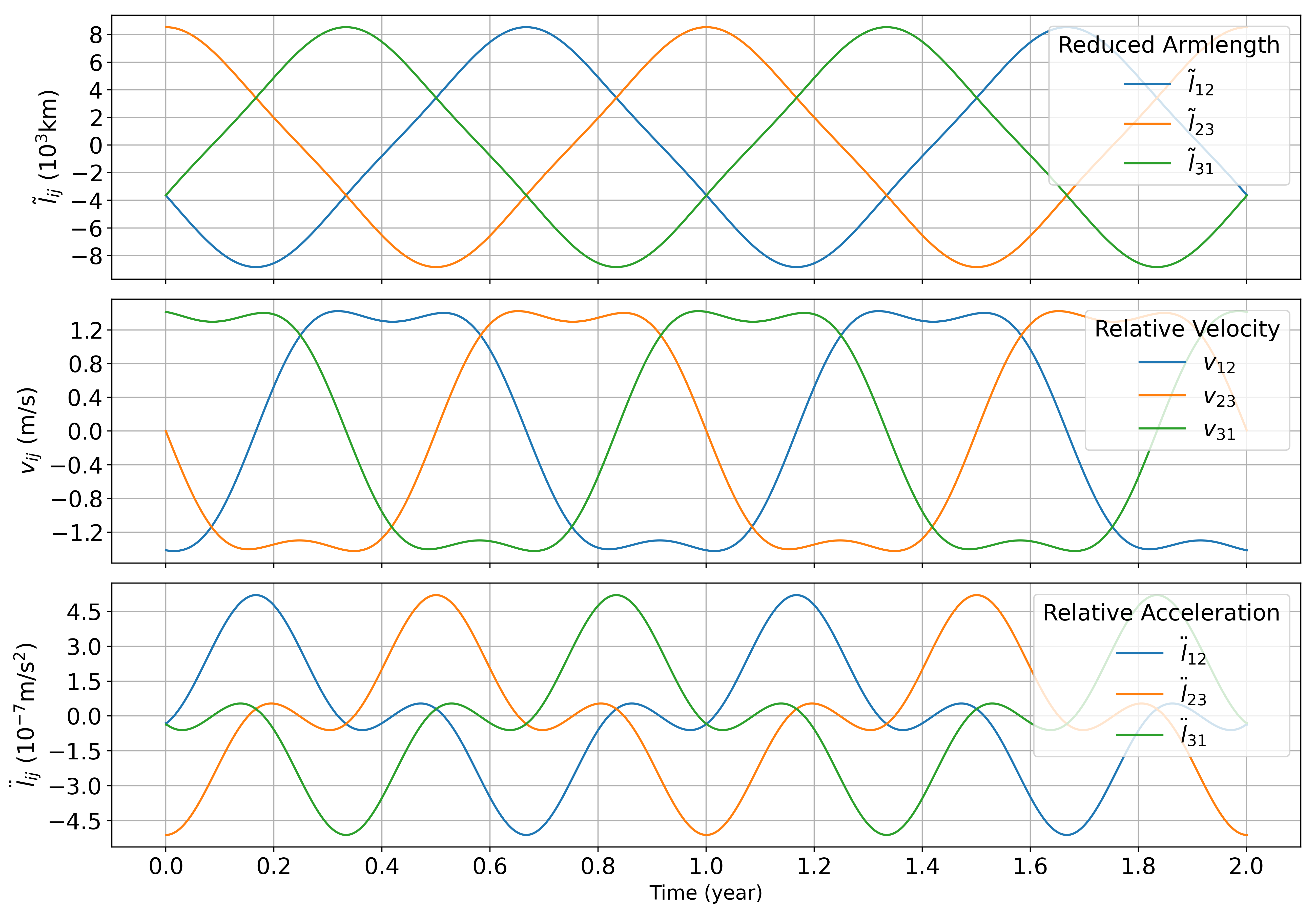}
		\caption{Armlength, rate of change in armlength, and relative acceleration in the Keplerian orbital configuration.}
		\label{Fig:armlength}
	\end{figure}

\begin{figure}[htbp]
		\centering
		\includegraphics[width=0.95\textwidth]{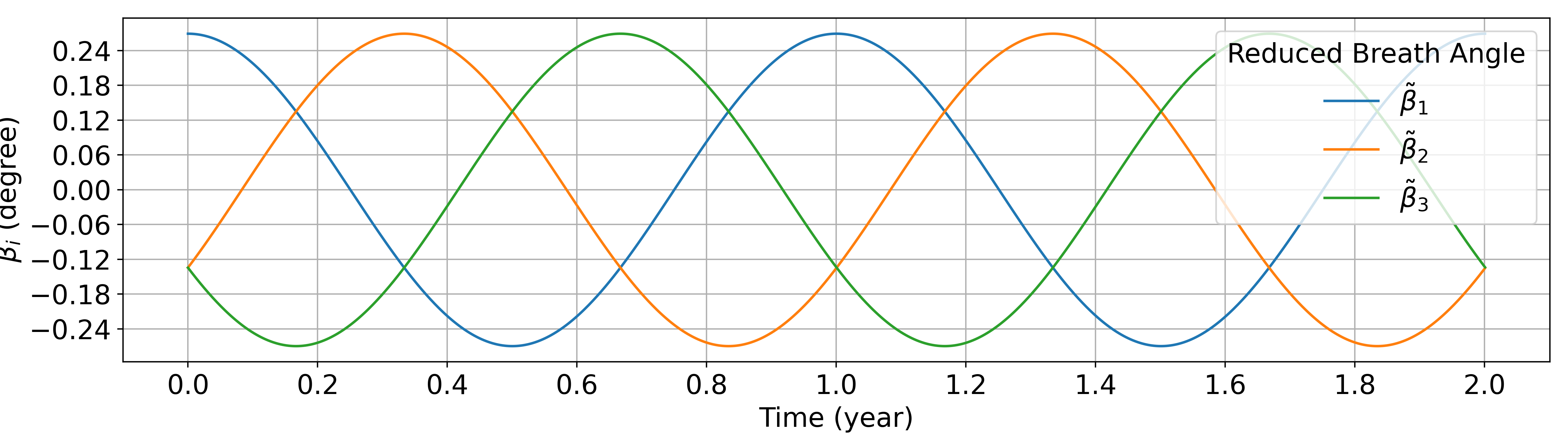}
		\caption{Breathing angle in the Keplerian orbital configuration. }
		\label{Fig: Breathing angle without perturbed}
	\end{figure}

	By defining the reduced armlength as $\tilde{l}_{ij} = l_{ij}-3\times10^6\,{\rm km}$ and the reduced breathing angle as $\tilde{\beta}_{i}=\beta_{i}-60^{\circ}$, we obtain the variation of Taiji constellation reduced armlengths, armlength variation rate, and reduced breathing angles within six years, as shown in Figures\,\ref{Fig:armlength} and \ref{Fig: Breathing angle without perturbed}.
	In the Keplerian orbital configuration, the reduced armlength amplitude is approximately $1\times10^{4}\,{\rm km}$, the rate of the armlength amplitude variation is $1.2\,{\rm m/s}$, breathing angle amplitude is $-0.24^{\circ}$$\sim$$0.24^{\circ}$, and the relative acceleration amplitude is approximately $4.5 \times10^{-7}\,{\rm m/s^2}$.

	\section{Taiji Heliocentric Formation Configuration}\label{sec3}
	
	\subsection{Influence of the Gravitational Field of the Solar System {Bodies}}
	As we aim to evaluate the influence of the gravitational fields of celestial bodies on the configuration of spacecraft formation, we consider only the Newtonian gravitational forces of the Sun and other $9$ bodies, Mercury, Venus, Earth, Mars, Jupiter, Saturn, Uranus, Neptune, and the Moon, in the solar system ($k = 1,2,...,9$). The motion equation of the spacecraft in heliocentric ECLIPJ2000 can be expressed as
	\begin{equation}\label{Equ: equation of motion}
		\boldsymbol{\ddot{r}}_i=-\mu_s\frac{\boldsymbol{r}_i}{|\boldsymbol{r}_i|^3}
		+\sum_k\mu_k\left(\frac{\boldsymbol{r}_k-\boldsymbol{r}_i}{|\boldsymbol{r}_k-\boldsymbol{r}_i|^3}-\frac{\boldsymbol{r}_k}{|\boldsymbol{r}_k|^3}\right),
	\end{equation}
	where $\boldsymbol{r}_i$ is the spacecraft position vector, $\mu_s = GM_s$, $\mu_k = GM_k$, $s$ denotes the Sun, and $G$ is the gravitational constant. 
	
	For this specific formation configuration, the initial mission time is 00:00:00 on 27 October 2032 (heliocentric ecliptic coordinate system), and the initial position and velocity are presented in Table\,\ref{table: Initial spacecraft Position and Velocity (October 27, 2032, 00:00:00)}.
	\begin{table}[htbp]
		\caption{Initial spacecraft position and velocity (00:00:00 on 27 October 2032).}
		\label{table: Initial spacecraft Position and Velocity (October 27, 2032, 00:00:00)} 
		\centering 
		\setlength{\tabcolsep}{2.5mm}
		\begin{tabular}{ccccccc}                   
			\toprule     &\boldmath{$x~(\mathrm{km})$}&\boldmath{$y~(\mathrm{km})$}&\boldmath{$z~(\mathrm{km})$}&\boldmath{$v_x~(\mathrm{km}\cdot\mathrm{s}^{-1})$}&\boldmath{$v_y~(\mathrm{km}\cdot\mathrm{s}^{-1})$}&\boldmath{$v_z~(\mathrm{km}\cdot\mathrm{s}^{-1})$}\\
			\midrule
			\text{SC1}& 89,327,505.9 & 119,295,324 & $-$964,610.427 & $-$24.0101710 & 17.8122817 & $-$0.228092013\\
			\text{SC2}& 89,361,219.1 & 121,025,909 &  1,482,827.40 & $-$23.8434160 & 17.5664204 & $-$0.0548688499\\
			\text{SC3}& 87,111,566.8 & 121,271,130 & $-$479,093.919 & $-$24.1388383 & 17.5416846 & 0.282942901\\
			\bottomrule
		\end{tabular}
	\end{table}
	
\begin{figure}[htbp]
		\centering		
		\includegraphics[width=0.63\textwidth]{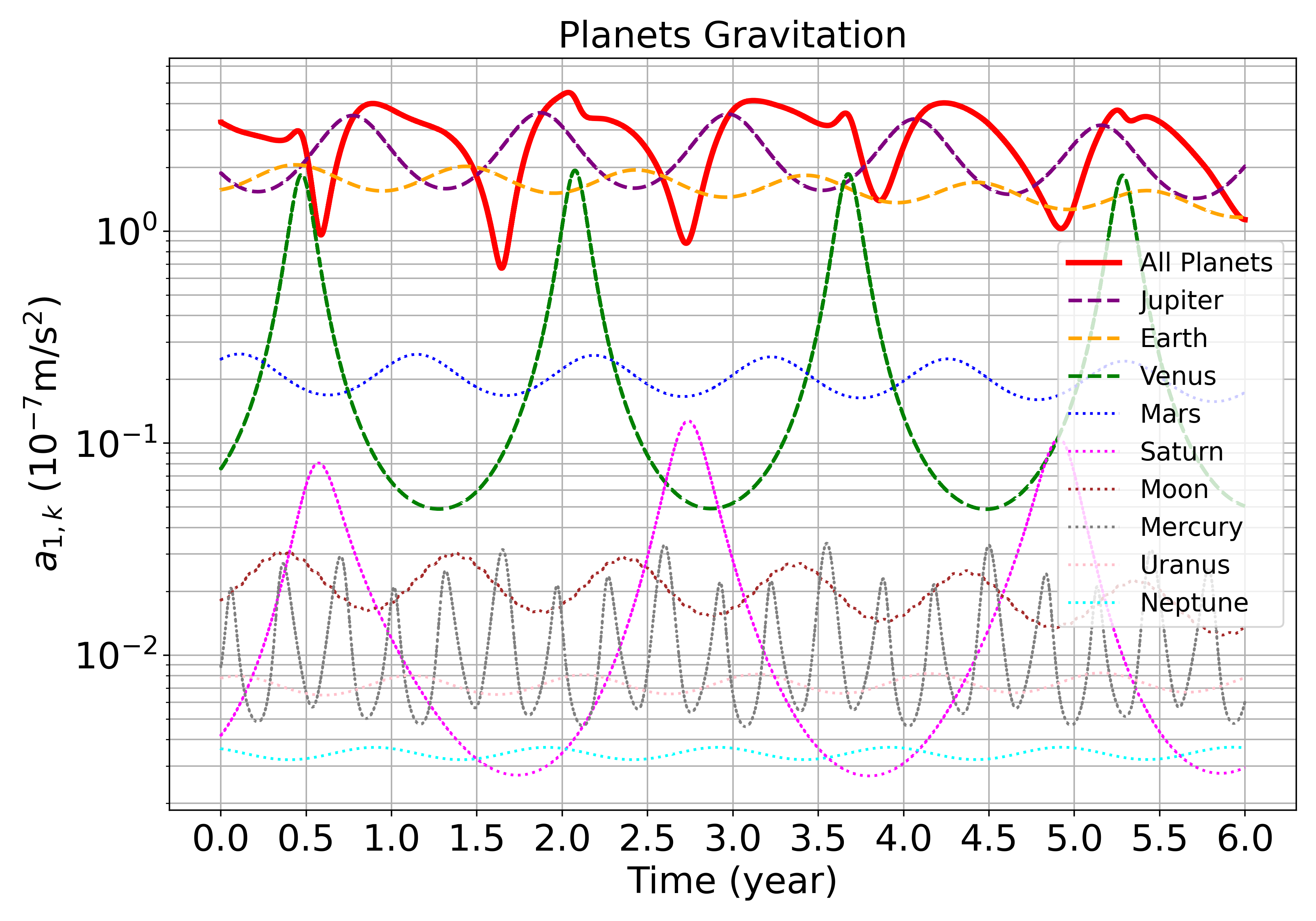}
		\caption{Acceleration of SC1 by the planets and the Moon in the solar system.}
		\label{Fig: Acceleration of SC1 by the planets and the Moon in the solar system.}
	\end{figure}

\begin{figure}[htbp]
		\centering
		\includegraphics[width=0.85\textwidth]{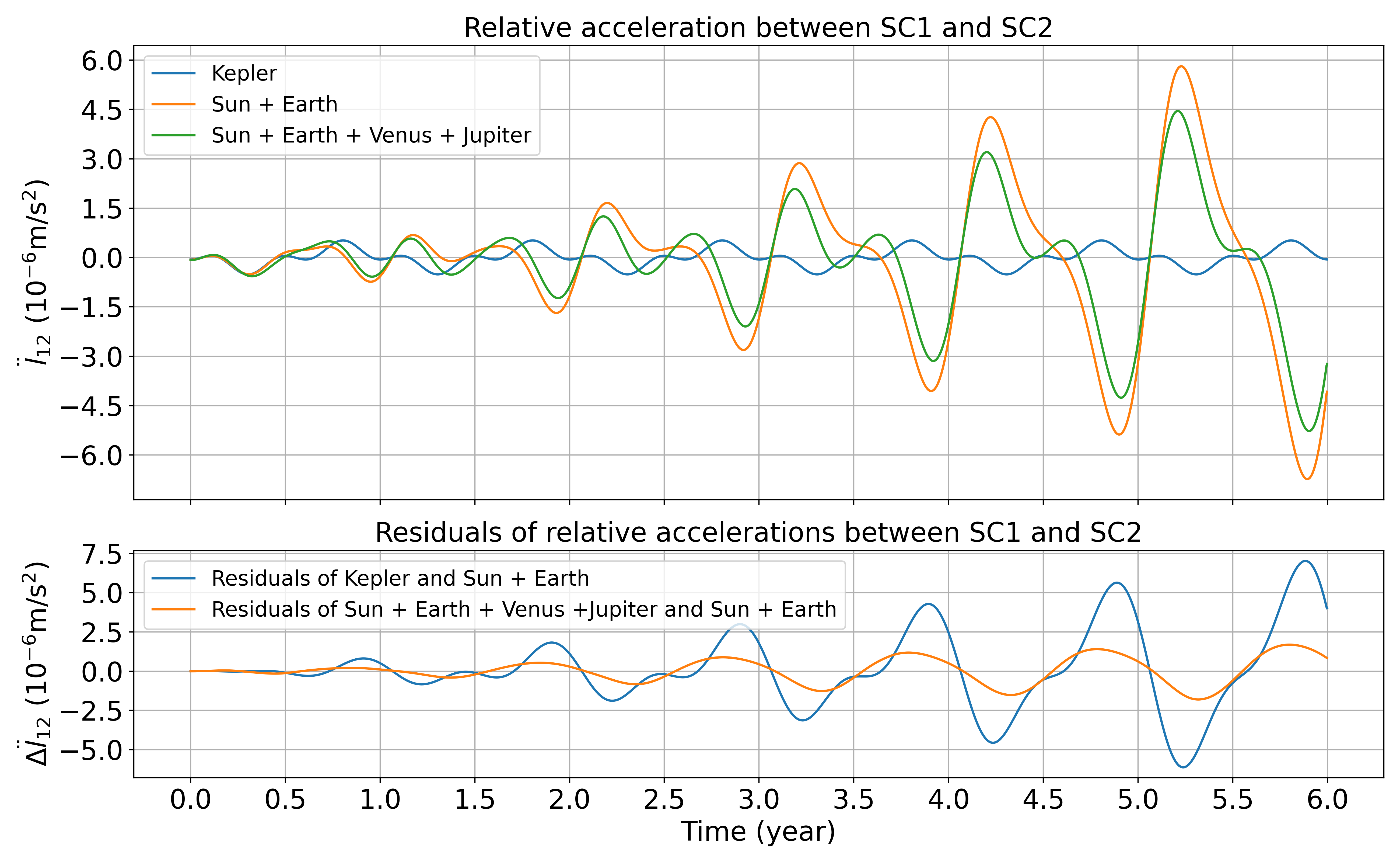}
		\caption{Relative acceleration between SC1 and SC2.}
		\label{Fig: Relative acceleration between SC1 and SC2}
	\end{figure}

	The acceleration of each planet toward SC1 is 
	\begin{equation}
		a_{i,k}=\mu_k\left|\frac{\boldsymbol{r}_k-\boldsymbol{r}_i}{|\boldsymbol{r}_k-\boldsymbol{r}_i|^3}\right|. 
	\end{equation}
	We obtain the planet position from the DE440 ephemeris, substitute it into the\linebreak Equation\,\eqref{Equ: equation of motion}, and obtain the result as shown in Figure\,\ref{Fig: Acceleration of SC1 by the planets and the Moon in the solar system.}.
	The Figure shows that the gravitational attractions exerted by the Earth, Venus, and Jupiter on spacecrafts are greater than those of other celestial bodies in the solar system and may have a greater variation on the spacecraft formation configuration. The contribution of the gravitational disturbances of the Earth, Venus, and Jupiter to the relative acceleration between spacecrafts obtained by numerical simulation is shown in Figure\,\ref{Fig: Relative acceleration between SC1 and SC2}.
	As shown in Figures\,\ref{Fig: Acceleration of SC1 by the planets and the Moon in the solar system.} and \ref{Fig: Relative acceleration between SC1 and SC2}, the gravitational perturbations of Venus, Earth, and Jupiter cause a variation in the spacecraft orbit. The maximum variation is of $\mathcal O(10^{-7})\,{\rm m/s^2}$. The Earth's gravitational perturbation has the greatest variation in the heliocentric formation configuration of $\mathcal O(10^{-6})\,{\rm m/s^2}$. Subsequently, we will examine variations in spacecraft formation configuration parameters perturbed solely by Earth's gravity.

	\subsection{Influence of the Gravitational Perturbation of the Earth}
	Considering the gravitational perturbation of the Earth, the motion equation of the spacecraft can be expressed as
	\begin{equation}\label{Equ: Earth perturbation acceleration}
		\boldsymbol{\ddot{r}}_i=-\mu_s\frac{\boldsymbol{r}_i}{|\boldsymbol{r}_i|^3}+\mu_3\left(\frac{\boldsymbol{r}_3-\boldsymbol{r}_i}{|\boldsymbol{r}_3-\boldsymbol{r}_i|^3}-\frac{\boldsymbol{r}_3}{|\boldsymbol{r}_3|^3}\right).
	\end{equation}
	Substituting the initial conditions in Table\,\ref{table: Initial spacecraft Position and Velocity (October 27, 2032, 00:00:00)} into Equation\,\eqref{Equ: Earth perturbation acceleration}, the spacecraft position vector $\boldsymbol{r}_i(t)$ can be obtained. Substituting $\boldsymbol{r}_i(t)$ into Equation\,\eqref{Equ:armlength, velocity, breathing angle, and relative acceleration} and comparing it with the Keplerian orbits, the result is shown in Figure\,\ref{Fig:armlength, breath angle and armlength rate changed under earth}.
	\begin{figure}[htbp]
		\centering
		\includegraphics[width=0.95\textwidth]{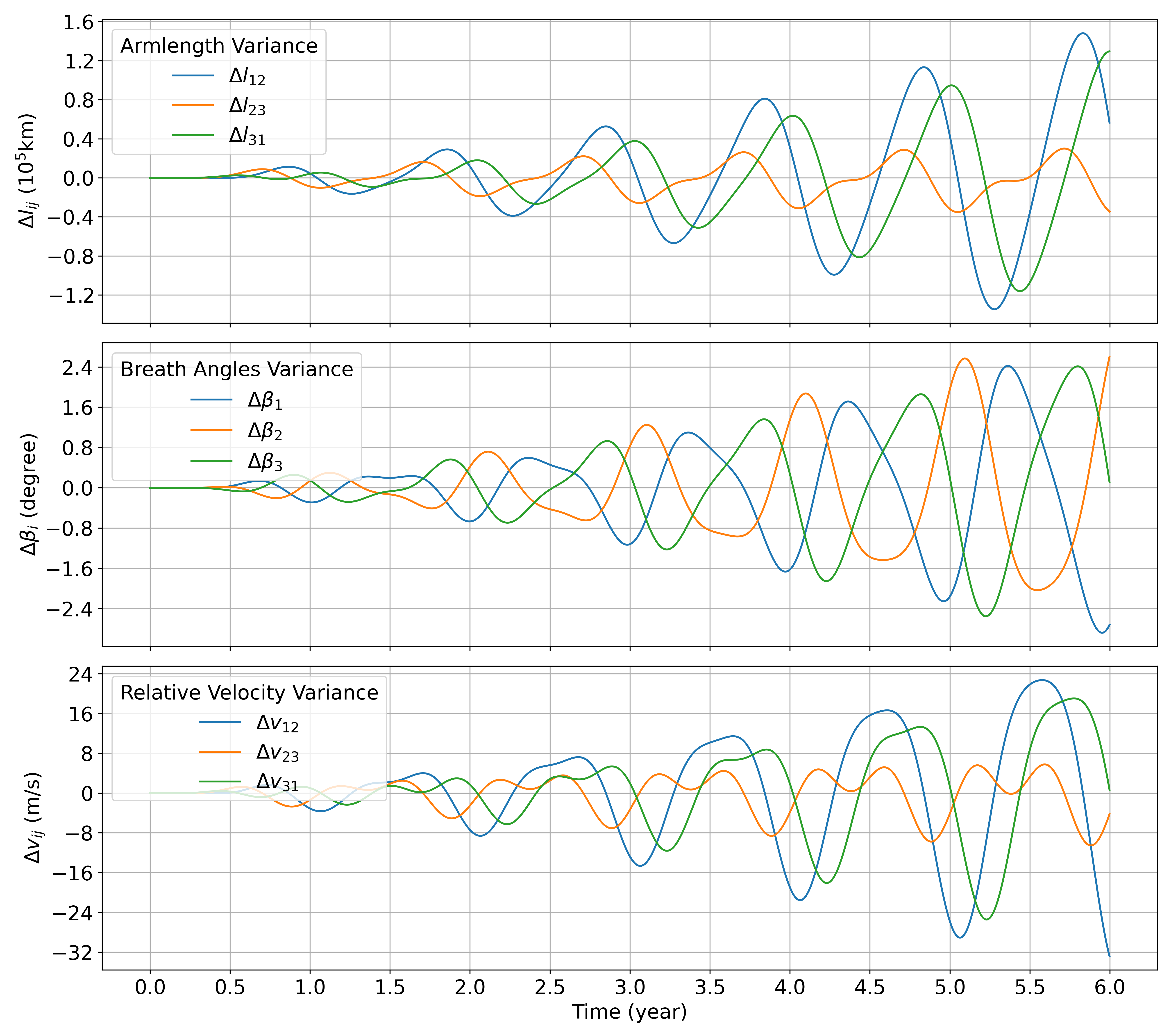}
		\caption{Considering the gravitational perturbation of the Earth, the variations in configuration parameters relative to the Kepler orbit within six years: (top) variation in the armlength, (middle) variation in the breathing angle, (bottom) variation in the armlength variation rate, when the orbital insertion time is 00:00:00 on 27 October 2032.}
		\label{Fig:armlength, breath angle and armlength rate changed under earth}
	\end{figure}
	
	Considering the Earth's gravitational perturbations, the orbits of spacecrafts in the heliocentric formation deviate from the Kepler orbit. This deviation changes the spacecraft formation configuration and increases the magnitude of the configuration parameters. Specifically, the maximum amplitude of the relative change in armlength between spacecrafts is $1.6\times10^5\,{\rm km}$, the maximum amplitude of the relative change in relative velocity between spacecrafts is $32\,{\rm m\cdot s ^{-1}}$, and the maximum relative change in the breathing angle is $3.2^{\circ}$ within six years.
	
	\subsection{Influence of the Gravitational Perturbation of Venus and Jupiter}
	To obtain accurate configuration parameter variations, the influence of the gravities of Venus and Jupiter on the spacecraft formation configuration must be considered. Choosing $k = 2, 3, 5$ to represent Venus, Earth, and Jupiter in the equation, respectively, for a single spacecraft in the heliocentric formation configuration, the following acceleration can be obtained.
	\begin{equation}\label{Equ: Earth Venus and Jupiter perturbation acceleration}
		\boldsymbol{\ddot{r}}_i=-\mu_s\frac{\boldsymbol{r}_i}{|\boldsymbol{r}_i|^3}+
		\mu_2\left(\frac{\boldsymbol{r}_2-\boldsymbol{r}_i}{|\boldsymbol{r}_2-\boldsymbol{r}_i|^3}-\frac{\boldsymbol{r}_2}{|\boldsymbol{r}_2|^3}\right)+
		\mu_3\left(\frac{\boldsymbol{r}_3-\boldsymbol{r}_i}{|\boldsymbol{r}_3-\boldsymbol{r}_i|^3}-\frac{\boldsymbol{r}_3}{|\boldsymbol{r}_3|^3}\right)+
		\mu_5\left(\frac{\boldsymbol{r}_5-\boldsymbol{r}_i}{|\boldsymbol{r}_5-\boldsymbol{r}_i|^3}-\frac{\boldsymbol{r}_5}{|\boldsymbol{r}_5|^3}\right).
	\end{equation}
	Substituting the initial conditions in Table.\,\ref{table: Initial spacecraft Position and Velocity (October 27, 2032, 00:00:00)} into Equation\,\eqref{Equ: Earth Venus and Jupiter perturbation acceleration}, the spacecraft position vector $\boldsymbol{r}_i(t)$ can be obtained.
	Substituting $\boldsymbol{r}_i(t)$ into Equation\,\eqref{Equ:armlength, velocity, breathing angle, and relative acceleration} and comparing it with the perturbation of the spacecraft formation configuration parameters caused exclusively by the Earth's gravity, the results are shown in Figure\,\ref{Fig:armlength, breath angle and armlength rate changed under venus and jupiter}. 
	\begin{figure}[htbp]
		\centering
		\includegraphics[width=0.95\textwidth]{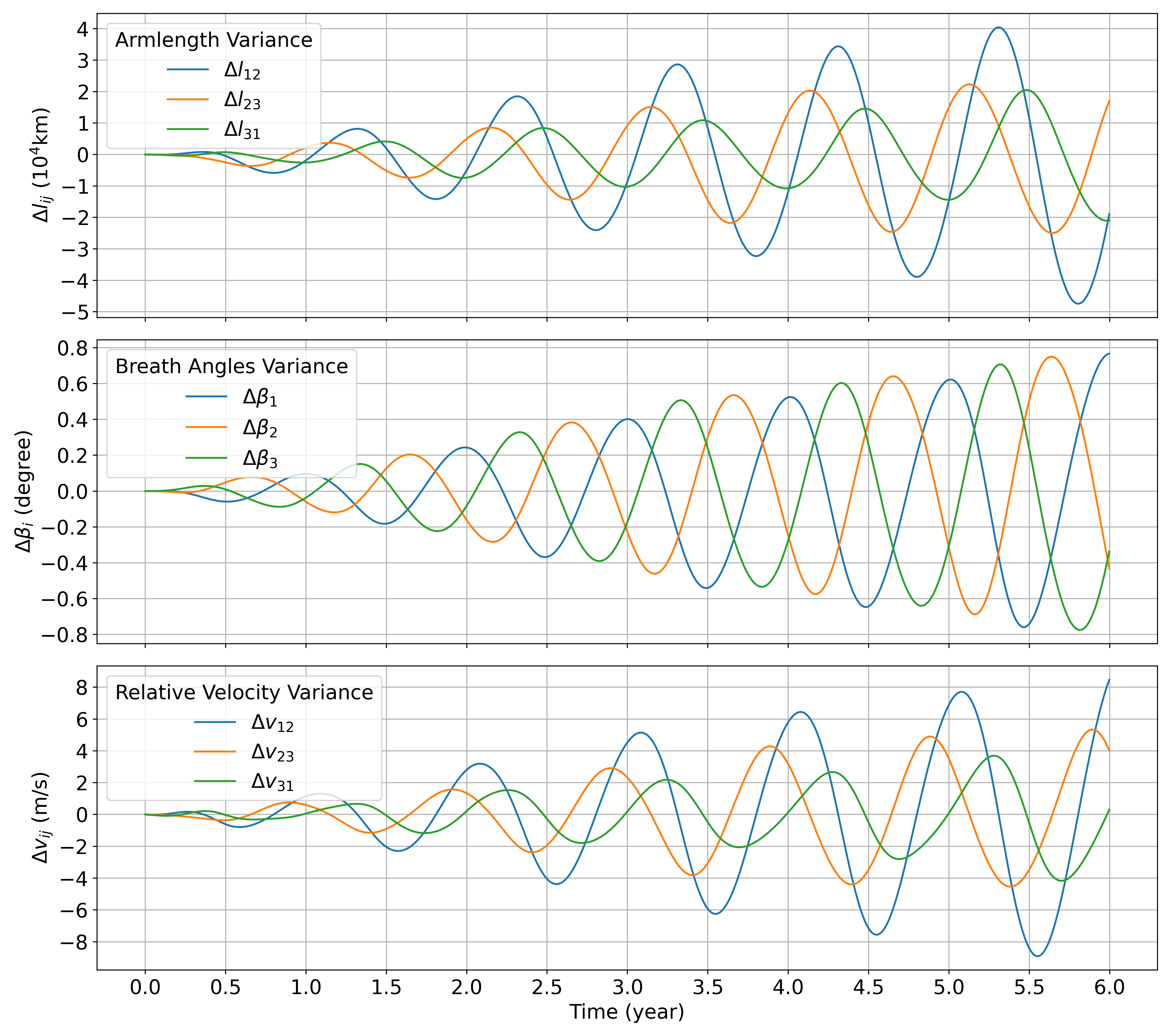}
		\caption{Considering the gravitational perturbations of Venus{, Earth} and Jupiter, the configuration parameters variations in $6$ years relative to the Earth-only perturbations: (top) change in armlength; (middle) change in breathing angle; and (bottom) change in armlength variation rate, when the orbital insertion time is 00:00:00 on 27 October 2032.}
		\label{Fig:armlength, breath angle and armlength rate changed under venus and jupiter}
	\end{figure}
	On can see that in addition to the gravity perturbation of the Earth, the configuration parameters will change when the gravity perturbation from Venus{, Earth}, and Jupiter is included. The maximum relative variations in inter-spacecraft armlength and relative velocity are $5\times 10^{4}\,{\rm km}$ and $9\,{\rm m\cdot s^{-1}}$, respectively, within six years. The maximum relative variation in breathing angle is approximately $0.8^{\circ}$.
	
\begin{figure}[htbp]
		\centering
		\includegraphics[width=0.95\textwidth]{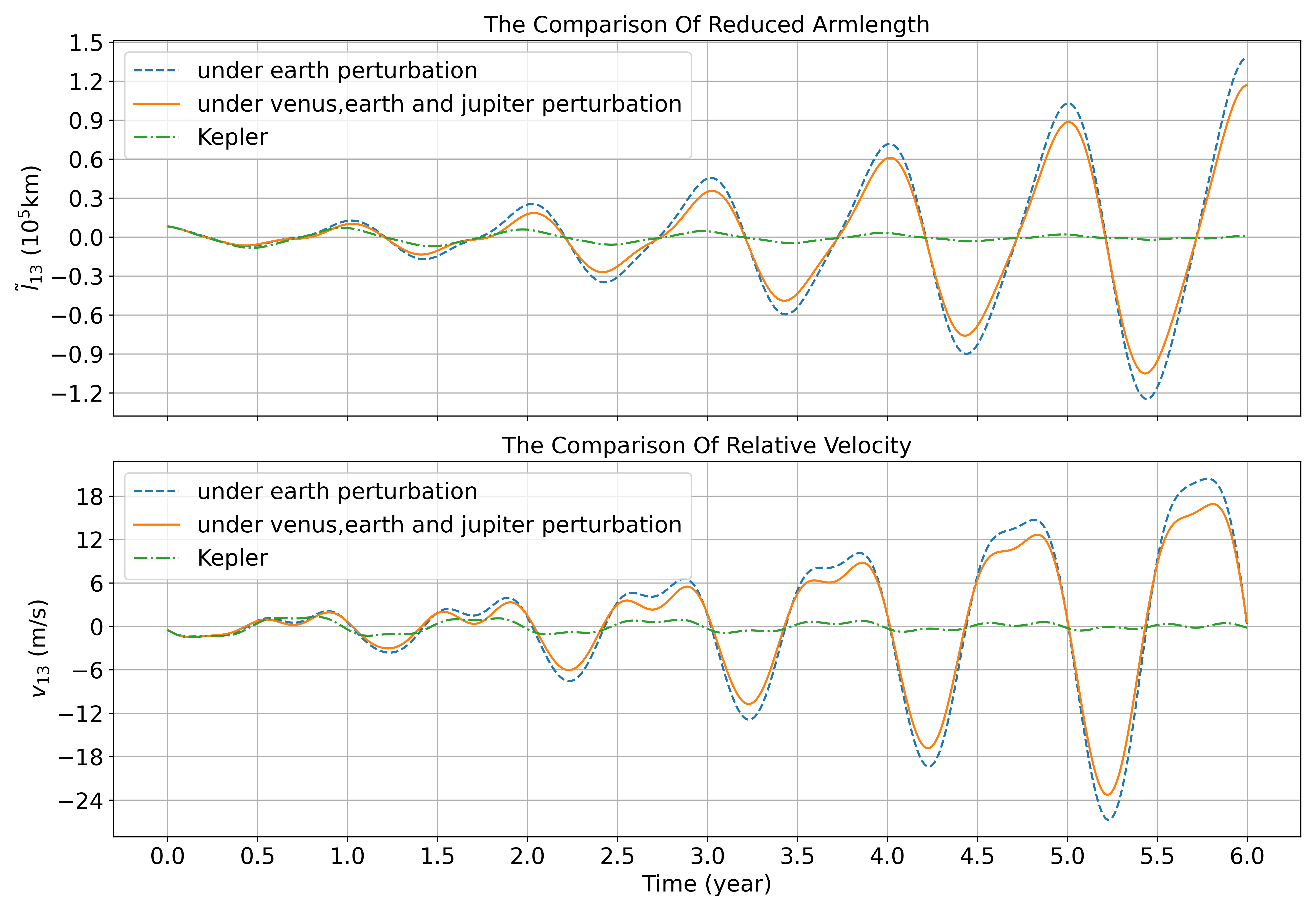}
		\caption{Comparison of spacecraft formation configuration parameters in three cases: (top) reduced armlength and (bottom) armlength variation rate.}
		\label{Fig: Comparison of spacecraft formation configuration parameters in three cases}
	\end{figure}

	However, the relative magnitude of the changes does not accurately reflect the influence of the perturbations of Venus and Jupiter. Figure\,\ref{Fig: Comparison of spacecraft formation configuration parameters in three cases} compares the changes in spacecraft formation configuration parameters in the three cases: the unperturbed Keplerian configuration, adding only the gravitational perturbation of the Earth, and adding the perturbations of Venus and Jupiter.
	As shown in Figure\,\ref{Fig: Comparison of spacecraft formation configuration parameters in three cases}, when considering the gravitational disturbances of Venus and Jupiter in addition to the gravitational disturbances of the Earth, the reduced armlengths and armlength variation rate decrease by $16.01\%$ and $17.45\%$, respectively.
	
	The findings indicate that when the orbit entry time is set to 00:00:00 on  27 October 2032, the Earth's gravity amplifies the variations in the constellation armlength, change rate, and breathing angle. However, the gravitational perturbations of Venus and Jupiter reduce the magnitude of the constellation configuration parameters.

	\subsection{Orbital entry moment}
\begin{figure}[htbp]
		\centering
		\includegraphics[width=0.9\textwidth]{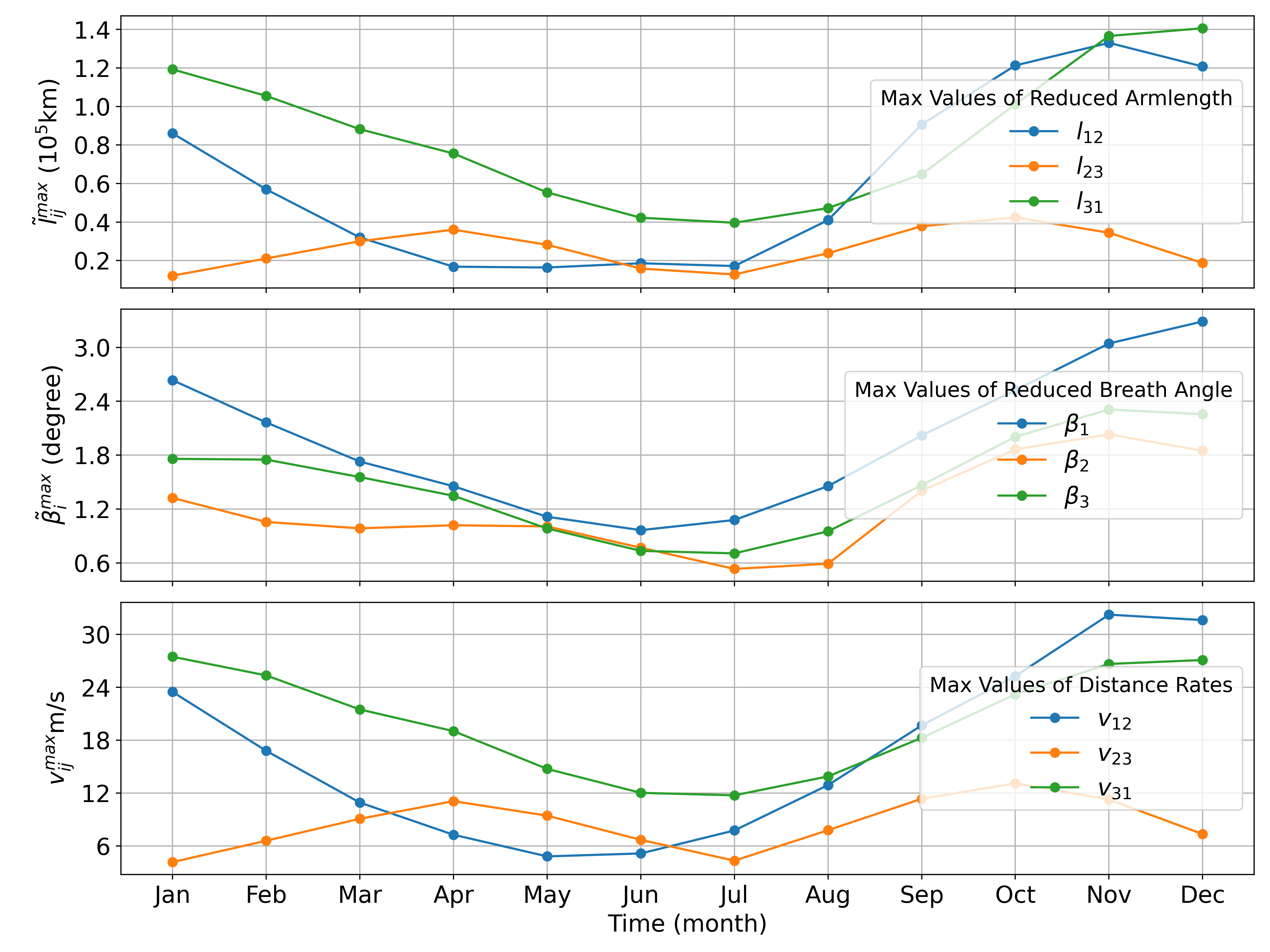}
		\caption{From top to bottom, the three rows represent the maximum amplitude of the reduced armlength, the maximum amplitude of the reduced breathing angle, and the maximum amplitude of the inter-spacecraft relative velocity.}
		\label{Fig: moment of orbit entry}
	\end{figure}

	The orbital entry moment is an optimization variation that affects the orbit design, and the different initial orbiting moments change the initial position and velocity of the spacecraft. In this section, different initial positions and velocities are obtained by changing the initial orbital entry moment of the formation. Further, the changes in configuration parameters at different orbital entry moments are analyzed. The specified orbiting time is set as 00:00:00 on the first day of each month in 2032. The simulation assesses the variations in the armlengths, breathing angles, and armlength variation rates over six years following the different orbiting times, which are shown in Figure\,\ref{Fig: moment of orbit entry}. 
	Notably, variations in the maximum amplitudes of the configuration parameters are observed at different orbital moments. Specifically, in July, the maximum amplitudes of the armlength and relative velocity between the spacecrafts are the smallest over one year. Consequently, selecting July as the moment of orbit entry is conducive to extending the mission duration.
	
\begin{figure}[htbp]
		\centering
		\includegraphics[width=0.95\textwidth]{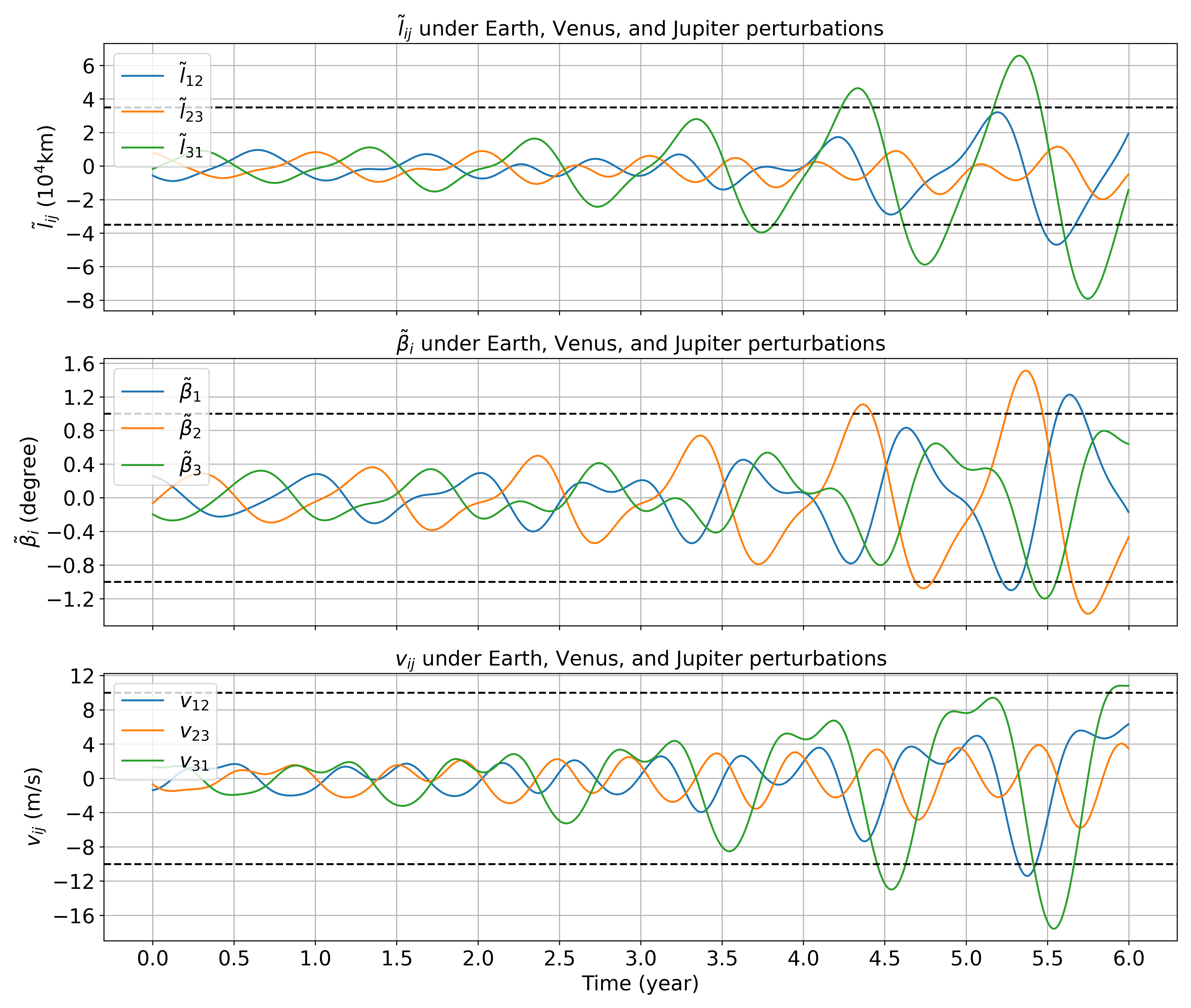}
		\caption{The variations of the reduced armlength, reduced breathing angle, and the inter-spacecraft relative velocity when the orbital insertion time is 00:00:00 on 1 July 2032.}
		\label{Fig: Final result 0701}
	\end{figure}

	Through sensitivity analysis of the initial orbital entry moment of the heliocentric formation, the orbital moment chosen was 00:00:00 on  1 July 2032. The initial positions and velocities of the spacecraft are presented in Table\,\ref{table: Initial spacecraft position and velocity (July 1, 2032, 00:00:00).}.
	\begin{table}[htbp]
		\caption{Initial spacecraft position and velocity (00:00:00 on 1 July 2032).}
		\label{table: Initial spacecraft position and velocity (July 1, 2032, 00:00:00).} 
		\centering		
		\setlength{\tabcolsep}{2mm}
		\begin{tabular}{ccccccc}                   
			\toprule      &\boldmath{$x~(\mathrm{km})$}&\boldmath{$y~(\mathrm{km})$}&\boldmath{$z~(\mathrm{km})$}&\boldmath{$v_x~(\mathrm{km}\cdot\mathrm{s}^{-1})$}&\boldmath{$v_y~(\mathrm{km}\cdot\mathrm{s}^{-1})$}&\boldmath{$v_z~(\mathrm{km}\cdot\mathrm{s}^{-1})$}\\
			\midrule    
			\text{SC1}& 73,144,893.1 & $-$131,441,026 & 1,448,987.09 & 25.8589315 & 14.4439795 & $-$0.0829325697\\
			\text{SC2}& 74,530,827.3 & $-$129,478,529 & $-$338,159.520 & 25.9306467 & 14.7317007 & 0.290937711\\
			\text{SC3}& 71,806,563.1 & $-$130,521,832 & $-$1,071,715.25 & 26.1457428 & 14.5228729 & $-$0.2080306030\\
			\bottomrule
		\end{tabular}
	\end{table}
	By substituting the parameters in Table\,\ref{table: Initial spacecraft position and velocity (July 1, 2032, 00:00:00).} into Equation\,\eqref{Equ: Earth Venus and Jupiter perturbation acceleration}, the position vector $\boldsymbol{r}_i(t)$ of the spacecraft can be determined. The vector in Equation\,\eqref{Equ: Kepler position} is used to calculate the variation in spacecraft formation configuration parameters, as shown in Figure\,\ref{Fig: Final result 0701}.
	The findings indicate that selecting the orbital entry moment as 00:00:00 on   1 July 2032, while accounting for the gravitational perturbation of Venus, Earth, and Jupiter, results in a maximum amplitude of approximately $8\times10^4\,{\rm km}$ in the reduced armlengths, a maximum armlength variation rates of $18.1\,{\rm m\cdot s^{-1}}$, and a maximum amplitude of approximately $1.6^{\rm \circ}$ in the reduced breathing angles over six years.

	\section{Simulation with Different Ephemeris}\label{sec4}
	
	The accuracy of the numerical simulation results depends on that of the planetary position data. Differences in planetary orbital positions may occur in various ephemerides due to variations in the chosen fitting data and the integration method used~\cite{Uncertainties_of_ephemeris}. These deviations can affect the numerical simulation of spacecraft configuration parameters. Therefore, the biases in numerical simulations caused by different ephemerides as input should be extensively investigated. 
	
	We obtained planetary position data from the ephemerides DE440, DE430 and DE421~\mbox{\cite{DE440,DE430,DE421}}. In the heliocentric equatorial coordinate system J2000, the maximum disparities in the coordinates of Venus, Earth, and Jupiter relative to the solar center of mass are presented in Table\,\ref{table:The maximum coordinate differences of Venus, Earth, and Jupiter relative to the center of the Sun in the ECLIPJ2000 coordinate system.} across various ephemerides.
	\begin{table}[htbp]
		\caption{Maximum coordinate differences of Venus, Earth, and Jupiter relative to the center of the Sun in the ECLIPJ2000 coordinate system. }
		\label{table:The maximum coordinate differences of Venus, Earth, and Jupiter relative to the center of the Sun in the ECLIPJ2000 coordinate system.}
		\centering
		\setlength{\tabcolsep}{2.6mm}
		\begin{tabular}{cccccccccc}
			\toprule
			&  &  \textbf{Venus}&  &  &  \textbf{Earth}&  &  &  \textbf{Jupiter}&  \\
			\midrule
			& \boldmath{$x~(\mathrm{m})$} &  \boldmath{$y~(\mathrm{m})$} & \boldmath{$z~(\mathrm{m})$} &   \boldmath{$x~(\mathrm{m})$} &  \boldmath{$y~(\mathrm{m})$} & \boldmath{$z~(\mathrm{m})$} &   \boldmath{$x~(\mathrm{m})$} &  \boldmath{$y~(\mathrm{m})$} & \boldmath{$z~(\mathrm{m})$}   \\
			
			DE440-DE430& 401 & 403 & 47  & 488 & 497 & 81 & 33,402 & 23,274 & 39,157 \\
			
			DE440-DE421& 116 & 107 & 167  & 48 & 53 & 217& 4093 & 3422 & 66,693 \\
			
			DE430-DE421& 308 & 297 & 175& 459 & 464 & 233 & 32,134 & 36,529 & 27,605 \\
			\bottomrule
		\end{tabular}
	\end{table}
	Comparatively, the maximum coordinates of the ephemerides of the three planets differ significantly: Venus and Earth differ at the 100 m level, while Jupiter differs at the 1\mbox{0 km~level}. 
	
\begin{figure}[htbp]
		\centering
		\includegraphics[width=\textwidth]{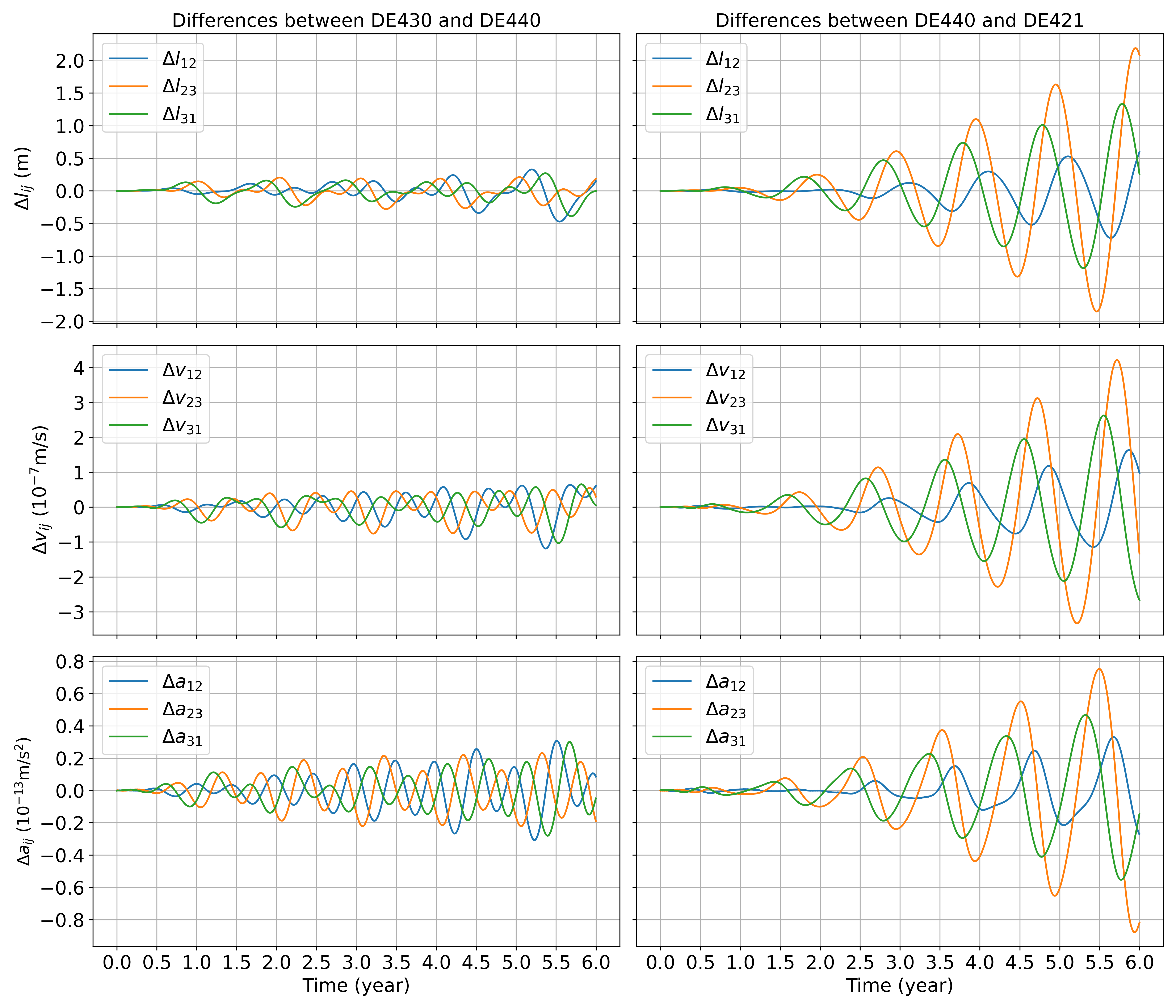}
		\caption{By using different ephemeris tables, the differences in the three motion indicators of the spacecraft formation are determined. The first line represents the differences in armlength, the second represents the differences in armlength variation rates between the spacecrafts, and the third represents the differences in relative acceleration between the spacecrafts. }
		\label{Fig:Differences generated by the ephemeris}
	\end{figure}

	Discrepancies in the planetary coordinates between ephemerides can cause variations in the forces acting on the spacecrafts. Consequently, these variations may alter the relative acceleration between the spacecrafts, thereby affecting the overall configuration of the formation. Three ephemerides, DE421, DE430, and DE440, were used for numerical simulations with an initial entry time of 00:00:00 on 1 July 2032. Figure\,\ref{Fig:Differences generated by the ephemeris} shows the difference in spacecraft armlengths, armlength variation rates, and relative accelerations within six~years. 
	
	Figure\,\ref{Fig:Differences generated by the ephemeris} shows differences in the spacecraft configuration parameters obtained via numerical simulation using different ephemerides. 
	Specifically, the variation between DE440 and DE430 is relatively small, with a difference of approximately $0.5\,{\rm m}$ in armlength over six years, $1\times10^{-7}\,{\rm m/s}$ in armlength variation rates, and $3\times10^{-14}\,{\rm m/s^2}$ in relative acceleration. Conversely, the variation between DE421 and DE440 is significant, with a difference of approximately $2.5\,{\rm m}$ in armlength, $5\times10^{-7}\,{\rm m/s}$ in armlength variation rates, and $1\times10^{-13}\,{\rm m/s^2}$ in relative acceleration within six years. 
	
	Notably, the variable period of the difference in spacecraft structural parameters obtained via the numerical simulation of different ephemerides is approximately one year, and the frequency is relatively low. This is relatively small compared to the change in the gravitational wave frequency band targeted by the Taiji project.

	\section{Conclusions}\label{sec5}
	
	Using the DE series of ephemerides, we investigated the influence of celestial gravitational perturbations on the heliocentric formation configuration.
	Specifically, variations in the major planets in the solar system on the orbital stability of the Taiji space gravitational wave detector were investigated. 
	Ephemeris DE440 was used to interpolate the planetary orbital position during the mission, and the high-order Runge--Kutta numerical integration method was used to calculate planetary gravity. 
	The trajectory of each spacecraft in the heliocentric formation configuration under the influence of perturbation was numerically simulated. In our investigation, to streamline the model, solar system bodies are regarded as point masses, with ancillary factors such as their geometric characteristics and rotational dynamics being omitted. These aspects will be meticulously scrutinized in forthcoming~research. 
	
	Results showed that the gravitational acceleration affecting an individual spacecraft within a spacecraft formation due to planetary gravity is approximately $10^{-7}\,{\rm m\cdot s^{-2}}$. 
	The relative acceleration between spacecrafts due to planetary gravity is approximately $10^{-6}\,{\rm m\cdot s^{-2}}$. 
	The difference in the heliocentric formation configuration is mainly affected by the gravity of the Earth. 
	Considering 00:00:00 on 27 October 2032 as the initial orbit entry moment as an example, when only the Earth's gravitational perturbation is considered, the maximum amplitude of the relative change in the armlength within six years was $1.6\times10^5\,{\rm km}$, the maximum amplitude of the relative change in breathing angle was $3.2^{\circ}$, and the maximum amplitude of the relative change in armlength variation rates was $32\,{\rm m\cdot s ^{-1}}$.
	
	Further, we discuss the spacecraft formation configuration changes with the addition of gravitational perturbations from Venus and Jupiter.
	The results showed that compared with the case where only the Earth's gravitational perturbation was considered, after the gravitational perturbations of Venus and Jupiter were added, within six years the maximum amplitude of the relative change in armlength was approximately $5 \times 10^{4}\,{\rm km}$, maximum amplitude of the breathing angle was approximately $0.8^{\circ}$, and maximum amplitude of the relative change in armlength variation rates was $9 \,{\rm m/s}$.
	After adding the gravitational perturbations of Venus and Jupiter, the armlength and relative velocity were reduced by $16.01\%$ and $17.45\%$, respectively, compared with when only the Earth's gravitational perturbation was considered.

	Variations in the maximum amplitudes of the formation configuration parameters were observed when entering orbit at different times. 
	The smallest increase in armlength variation rates between the spacecrafts occurred when the initial orbiting time was in July. Consequently, selecting July as the initial orbital insertion time extends the experiment duration.
	By considering 00:00:00 on 1 July  2032 as the initial orbit entry moment, under the gravitational perturbations of Venus, Earth, and Jupiter, using the DE440 ephemeris, the maximum amplitude of the reduced armlength was $8\times10^4\,{\rm km}$, the maximum amplitude of the reduced breathing angle was $1.6^{\circ}$, and the maximum armlength variation rates was $18.1\,{\rm m\cdot s^{-1}}$ within six years.
	
	Ephemeris data were used to perform numerical simulations. By comparison, the maximum disparity in the orbital positions of Venus and Jupiter was $300\,{\rm m}$ and $60\,{\rm km}$, respectively.
	The armlength deviation over the six-year simulation period was approximately $1\,{\rm m}$. 
	The deviation in armlength variation rates was in the order of $\mathcal O(10^{-7})\,{\rm m\cdot s^{-1}}$, and the deviation in relative accelerations was approximately $1\times10^{-13}\,{\rm m\cdot s^{-2}}$. 
	The differences between ephemerides DE440 and DE430 are smaller than those between DE440 and DE421. 
	Nonetheless, the impact of discrepancies in high-frequency bands induced by factors such as planetary rotation remains unexplored and warrants consideration in future studies.

\end{document}